\documentclass[a4paper]{aa}
\usepackage[T1]{fontenc}
\usepackage{graphicx}
\usepackage{booktabs}
\usepackage{hyperref}
\hypersetup{hidelinks}
\usepackage{pdflscape}

\newcommand{\erg}{~\mathrm{erg~s}^{-1}}
\newcommand{\keV}{~\mathrm{keV}}
\newcommand{\NH}{N_{\mathrm{H}}}
\newcommand{\chandra}{\textit{Chandra}}
\newcommand{\xspec}{\textsc{xspec}}

\begin{document}

\title{X-ray point sources towards the giant radio lobes of Centaurus~A}

\author{K.~Balasubramaniam\inst{1}
   \and H.-K.~Chang\inst{2}
   \and {\L}.~Stawarz\inst{3}
   \and V.~Marchenko\inst{3}}

\titlerunning{X-ray point sources towards the Cen~A giant radio lobes}
\authorrunning{Balasubramaniam et al.}

\institute{
  RAD@home Astronomy Collaboratory, Bengaluru, India\\
  \email{karthikedu77@gmail.com}
  \and
  Institute of Astronomy, National Tsing Hua University,
  101 Section~2 Kuang-Fu Rd., Hsinchu 30013, Taiwan
  \and
  Astronomical Observatory, Jagiellonian University,
  ul.~Orla~171, 30-244 Krak\'{o}w, Poland
}

\date{Received ...; accepted ...}

\abstract
{
Centaurus~A (NGC~5128) is the nearest radio galaxy ($d=3.8$~Mpc), so
individual X-ray point sources can be resolved against its giant radio
lobes.
}
{
We test whether a compact X-ray source population over the lobes exceeds
the expected field population once the selection function, foreground
stars and the cosmic X-ray background (CXB) are accounted for, and what
this implies for contamination of the diffuse lobe emission.
}
{
We re-process five \chandra/ACIS-I lobe pointings into a merged
$0.5$--$7.0\keV$ catalogue of 185 sources. Hardness ratios come from
background-marginalised intensity posteriors; sources are cross-matched
with Gaia~DR3 and AllWISE, and the eight brightest are fitted with an
absorbed power law. The selection function follows from the
\texttt{wavdetect} false-alarm probability and includes the catalogue
quality cut, so observed and expected counts are compared over identical
selections.
}
{
The hardness-ratio distribution (median $-0.17$, i.e.\
$\NH\approx1.7\times10^{21}\,$cm$^{-2}$; $80\%$ of sources below
$10^{22}\,$cm$^{-2}$) is that of background AGN plus a few foreground
stars. Above the flux at which the selection function covers half the
footprint, the good sample holds $N_{\rm obs}=83$ sources, 9 of them
foreground stars, against $N_{\rm exp}=101$ expected background sources.
The observed count therefore lies $27\%$ \emph{below} the expectation
($-1.5\sigma$ including the $15\%$ count-model normalisation
uncertainty), with no excess. The $90\%$ upper limit on a
lobe-associated excess above $6.9\times10^{36}\erg$ is
$N_{\rm lobe}<18$, or $<38$ for a conservative $30\%$ normalisation
uncertainty: a limit on the detected excess, not on an intrinsic
population.
}
{
These fields contain no point-source excess over the field expectation.
Converting the soft-band limit into stellar-mass or star-formation-rate
limits would require a spatial and luminosity-function model, so we do
not do so. With the usual resolved-source masking and blank-field CXB
subtraction, no additional lobe-associated correction to diffuse-emission
measurements, including inverse-Compton, is indicated. The catalogue is provided as arXiv ancillary material.
}


\maketitle

\section{Introduction}
\label{sec:intro}

The proximity of Centaurus~A (Cen~A; NGC~5128) at $d = 3.8 \pm 0.1$~Mpc \citep{Harris2010} makes it uniquely suitable among Fanaroff--Riley class~I radio galaxies \citep{Fanaroff1974}, and more generally among the population of nearby radio galaxies \citep{Tadhunter16}, for studying in detail the large-scale structure shaped by multiple episodes of central engine activity within the last Gyr \citep{Israel98,Morganti99,Eilek14,McKinley2022}. The most prominent features of Cen~A are the giant radio lobes extending north and south to distances of hundreds of kpc beyond the host galaxy \citep[see][]{Feain11,Stefan13,McKinley13}. The lobes have also been resolved at higher frequencies, including the mm-wavelength range, as well as in the high-energy $\gamma$-ray range, opening the discussion on particle acceleration within the giant relic radio structure, including the possibility that the lobes are sources of ultra-high-energy cosmic rays \citep{Hardcastle2009,Abdo2010,Yang2012,Wykes2013}. Indeed, the giant lobes in Cen~A span $\approx 600$~kpc end-to-end and record the integrated mechanical output of the AGN over ${\sim}\,5\times10^{8}$~yr, with a total mechanical energy deposited of ${\sim}\,10^{58}$--$10^{59}$~erg \citep{Abdo2010,Wykes2013,McKinley2022}, which is comparable to the binding energy of the hot halo of NGC~5128 \citep{Kraft2003}.

Due to their large angular extent ($\sim 5^{\circ} \times 9^{\circ}$) and low surface brightness, the giant lobes of Cen~A could hardly be probed in a meaningful way with the previous generation of X-ray instruments, such as SAS~3 or ROSAT, keeping in mind the intense emission of the active nucleus at X-ray frequencies, as well as the X-ray foreground/background emission varying on the scale of the giant lobes. In addition, the outer region of the Northern giant lobe was targeted by ASCA \citep{Isobe01}, while multiple selected regions partly overlapping with the complex system of highly polarized filaments in the Southern giant lobe \citep{Feain11,Wykes2014} were observed with Suzaku \citep{Stawarz2013} and XMM-Newton \citep{Wykes15}.

In particular, the detailed analysis of the Suzaku data reported in \citet{Stawarz2013} revealed the presence of multiple point-like sources at the angular resolution of a few arcmin provided by Suzaku's X-ray Imaging Spectrometer (XIS), some of which were concluded to be ``likely associated'' with the fine structure of the lobe seen in radio intensity and polarization maps. Even more interestingly, the Suzaku data spectroscopy and modeling tentatively revealed the presence of hot diffuse gas with a temperature of $0.5$\,keV within the lobes, in rough pressure balance with the non-thermal plasma emitting radio and $\gamma$-ray photons, as well as with the volume-averaged magnetic pressure in the lobes. This finding, although controversial to some extent, was corroborated by a detailed Rotation Measure study at radio wavelengths by \citet{OSullivan13}, which accordingly revealed a large amount of thermal gas, $\sim 10^{10}M_{\odot}$, mixed in with the non-thermal plasma within the giant lobes of Cen~A.

In the analysis of data such as the Suzaku pointings toward the Southern lobe of Cen~A, the main problems arise from limited photon statistics, the limited angular resolution of X-ray instruments, the expected presence of multiple emission components at X-ray wavelengths, and the issue of foreground/background emission, which may vary on degree scales and, at the same time, can only be estimated using off-pointings, since the source itself, i.e., the diffuse giant lobes, fills the entire field of view of the targeting instrument. These limitations render the resulting claims and findings tentative. Regarding the multiple emission components, one should in fact note that, in addition to the Cosmic X-ray Background (CXB), Galactic foreground X-ray emission, the gaseous environment of the Cen~A galaxy group surrounding the giant lobes, and the aforementioned potential thermal gas emission from within the lobes, the lobes themselves are expected to emit non-thermal X-ray photons due to inverse-Compton (IC) scattering of the CMB photon field by the radio-emitting electrons in the lobes \citep{Hardcastle2009,Yang2012,Wykes15}. Moreover, for a source with an angular extent of $\sim 5^{\circ} \times 9^{\circ}$, contamination by unrelated point sources, including background galaxies and AGN as well as foreground stars and X-ray binaries, may potentially become problematic.

In order to address the latter issue, as well as to investigate in more detail a possible distinct population of compact X-ray sources related to the radio structure of the Cen~A giant lobes, five \chandra{}/ACIS-I observations were proposed. The primary goal was to resolve, at arcsec resolution, individual X-ray point sources projected against the Cen~A giant radio lobes, with two pointings overlapping the Suzaku pointings to the south and three overlapping the ASCA pointings to the north. Here we analyze the resulting data, asking, in particular, whether, after accounting for the survey selection function and the CXB, any compact X-ray source population is physically associated with the giant lobes of Cen~A, and what the answer implies for constraints on the lobes' magnetic field derived from the available upper limits on their IC X-ray emission. Using these five dedicated \chandra{}/ACIS-I pointings, we build a vetted catalogue of 185 sources, characterize it using hardness ratios, bright-source spectroscopy, Gaia~DR3 and AllWISE associations, and completeness-corrected source counts, and use the resulting background model and sky-coverage function to place a quantitative upper limit on a lobe-associated excess and to constrain its effect on the IC-derived magnetic field.

We note in this context that \chandra{} observations of NGC~5128, the host galaxy of Cen~A, have established a rich X-ray point-source population. \citet{Woodley2008} detected 353 X-ray point sources (predominantly LMXBs) in four \chandra{} observations. \citet{Voss2009} derived the LMXB luminosity function and investigated its dependence on globular cluster membership. \citet{Burke2013} identified three transient black-hole LMXB candidates coincident with the dust lane, while \citet{Sivakoff2008} reported a transient black-hole low-mass X-ray binary candidate based on \chandra{} monitoring. A sub-population of supersoft sources and background AGN was characterised by \citet{Kraft2001} and \citet{Prestwich2003}. Such low-mass X-ray binary populations trace the integrated stellar mass of their host galaxies and follow well-characterised luminosity functions \citep{Gilfanov2004,Kim2004,Fabbiano2006}, while high-mass X-ray binaries track recent star formation \citep{Grimm2003,Mineo2012}; neither population is expected to contribute significantly at the large projected radii of the giant lobes, far beyond the main stellar body of the galaxy. The Cen~A jet and inner lobes are themselves bright X-ray structures \citep{Feigelson1981,Burns1983,Croston2009,Neff2015}, but these studies concentrate on the inner ${\sim}10'$ of the galaxy; X-ray point sources projected onto the giant lobes at degree-scale radii have not been the subject of a dedicated census.

Throughout we adopt the direct distance of \citet{Harris2010} ($3.8$~Mpc); at this distance $1'' = 18.4$~pc. All uncertainties are at $90\%$ confidence unless noted otherwise.

\section{Observations and data reduction}
\label{sec:obs}

\subsection{Dataset}
\label{sec:dataset}

We analyse five \chandra/ACIS-I pointings obtained in Cycle 14 (PI: Y. Tanaka; Proposal 14900703; ObsIDs 15199--15203) that were placed on the northern and southern giant lobes of Cen~A, two on the southern lobe (fields S1, S2) and three on the northern lobe (N1, N2, N3).  Each pointing has a clean exposure of $\approx20$~ks and lies $\gtrsim2^{\circ}$ from the nucleus, well outside the LMXB-dominated inner galaxy.  The observation log is given in Table~\ref{tab:obslog}.

\begin{table*}
\centering
\caption{\textit{Chandra}/ACIS-I observations of the Centaurus~A giant-lobe fields, all from 2013 (programme \emph{Studying Giant Filamentary Lobes of Centaurus~A}, PI: Y.~T.~Tanaka). $N_{\rm src}$ is the number of vetted point sources detected in each field.}
\label{tab:obslog}
\begin{tabular}{lccccccc}
\toprule
ObsID & Field & Lobe & R.A. & Dec. & Date & Exp. & $N_{\rm src}$ \\
 & & & (deg) & (deg) & & (ks) & \\
\midrule
15199 & S1 & S & 200.8668 & -45.1637 & 2013-04-15 & 19.5 & 29 \\
15200 & S2 & S & 200.3965 & -45.1399 & 2013-04-15 & 19.5 & 36 \\
15201 & N1 & N & 201.4693 & -38.9377 & 2013-04-08 & 19.8 & 47 \\
15202 & N2 & N & 201.7853 & -39.7630 & 2013-04-08 & 19.2 & 30 \\
15203 & N3 & N & 201.1589 & -39.3103 & 2013-05-01 & 19.8 & 43 \\
\midrule
\multicolumn{7}{l}{\textit{Total}} & 185 \\
\bottomrule
\end{tabular}
\end{table*}

\subsection{Data reduction}
\label{sec:reduction}

Event files were reprocessed with \textsc{CIAO}~4.18
\citep{Fruscione2006} and \textsc{CALDB}~4.12.4 using
\texttt{chandra\_repro}.  Full-field background light curves
($0.5$--$7\keV$, $259.28$~s bins, point sources masked) were extracted
and cleaned with \texttt{deflare} (\texttt{lc\_sigma\_clip},
$3\sigma$); no significant background flares were found, and the
exposures in Table~\ref{tab:obslog} are the post-filtering good time.
Absolute astrometry was registered against Gaia~DR3
\citep{GaiaMission2016,GaiaDR3} using \texttt{wcs\_match} on the
brightest sources in each field.  The measured field offsets
($0.09''$/$0.37''$/$0.17''$ from $6$/$5$/$3$ matches in N1/S1/S2) are
statistically consistent with zero at the ACIS pixel scale
($0.492''$), so no correction was applied; the two northern fields N2
and N3 had a single Gaia match each ($2.15''$, $0.53''$), too few for a
reliable solution.

\subsection{Source detection and photometry}
\label{sec:detection}

Point-source detection was performed per field with \texttt{wavdetect}
\citep{Freeman2002} on images in four standard \textsc{CIAO} energy
bands --- broad ($0.5$--$7\keV$), soft ($0.5$--$1.2\keV$), medium
($1.2$--$2.0\keV$) and hard ($2.0$--$7\keV$) --- at wavelet scales
$(1,2,4,8,16)$~px and a per-pixel detection threshold
\texttt{sigthresh}$=10^{-6}$ (nominally one false pixel per $10^{6}$,
i.e.\ $\lesssim1$ spurious detection per band-image; the multi-band,
multi-scale search raises the effective false rate, which the
photometric confirmation below removes).  Per-band detections were
merged within $2''$, retaining the highest-significance detection, and
a cut on the \texttt{wavdetect} output significance
(\texttt{SRC\_SIGNIFICANCE}$\geq3.3$, chosen empirically to suppress the
low-significance wavelet-peak tail while retaining real sources) was
applied, giving 221 merged candidates.  Aperture photometry with \texttt{srcflux} (PSF aperture
correction via \texttt{arfcorr}) then confirmed each source: 36
detections with zero \emph{gross} broad-band counts in the source
aperture --- wavelet noise peaks, concentrated in the off-axis northern
fields --- were rejected as spurious.  This photometric confirmation
yields a final catalogue of 185 sources (Table~\ref{tab:obslog}); a few
confirmed sources have zero \emph{net} broad-band counts after
background subtraction and are flagged as marginal
(Sect.~\ref{sec:catalogue}).  The full detection-to-catalogue sample
flow is given in Table~\ref{tab:sampleflow}.

\begin{table}
\centering
\footnotesize
\caption{Sample flow from raw detection to the counts analysis.  The last block is the matched sample: the flux interval, area, quality cut and star removal used on both sides of the comparison in Sect.~\ref{sec:constraints}.}
\label{tab:sampleflow}
\begin{tabular}{@{}p{0.80\linewidth}r@{}}
\toprule
Stage & Number \\
\midrule
\texttt{wavdetect} detections (4 bands, merged $<2''$) & 267 \\
After \texttt{SRC\_SIGNIFICANCE}\,$\geq3.3$ cut & 221 \\
After photometric confirmation (gross broad ct $\geq1$) & \textbf{185} \\
\quad ``good'' (broad $N_{\rm net}\geq5$ and signif $\geq3$) & 114 \\
\quad with $F_{0.5-2}>0$ & 113 \\
\midrule
\quad above $S_{\rm ref}$ (matched sample) & \textbf{83} \\
\qquad foreground stars & 9 \\
\qquad AGN-candidate detections & 74 \\
\bottomrule
\end{tabular}
\end{table}

Net count rates in the soft ($S$: $0.5$--$2.0\keV$) and hard ($H$:
$2.0$--$7.0\keV$) bands were measured with \texttt{srcflux} and
converted to fluxes assuming an absorbed power law with $\Gamma = 1.7$
and $\NH = 8.4\times10^{20}$~cm$^{-2}$ (Galactic value toward Cen~A;
\citealt{Kalberla2005}), with \citet{Wilms2000} abundances.
Because background subtraction can drive a band's net counts negative at
low count levels --- leaving the naive net-count ratio ill-defined and
formally outside $[-1,+1]$ --- the hardness ratio is evaluated from the
posterior distributions of the band intensities rather than from the net
counts, following the bounded-estimator approach of \citet{Park2006}.
For each band the counts in the source and background apertures are
modelled as independent Poisson variables whose means both contain a
source and a background contribution, with Jeffreys priors on the
aperture source intensity $\lambda$ and on the background surface
density; the background is then \emph{marginalised over} the observed
background-aperture counts rather than fixed at its gross-minus-net
value, and the source's own leakage into the background aperture is
retained.  In the presence of background the posterior for $\lambda$ is
not a gamma density but a finite mixture of gamma densities
\citep{Primini2014}, so we evaluate it numerically.
Appendix~\ref{app:hr} gives the photometric model, priors, estimator,
validation against the closed-form posterior of \citet{Primini2014}, and
the effect of each modelling choice on HR.  $\mathrm{HR}$ is the
posterior median, with the $68\%$ highest-posterior-density (HPD)
interval as its uncertainty.  The HPD interval is narrower than the
equal-tail interval for all 185 sources, but only slightly: the median
half-width is $0.20$ for the good sources either way and $0.33$ (instead
of $0.35$) for the marginal ones.  We do \emph{not} use the posterior
mode: at the $1$--$5$ counts typical of marginal sources the HR posterior
is monotonic and the mode migrates to $\pm1$, differing from the median
by more than $0.1$ for $49\%$ of the marginal sources against $2\%$ of
the good ones.  HR thus stays within $[-1,+1]$ for every source, at the
cost of large error bars on the lowest-count detections.

\begin{figure*}
  \centering
  \includegraphics[width=0.92\textwidth]{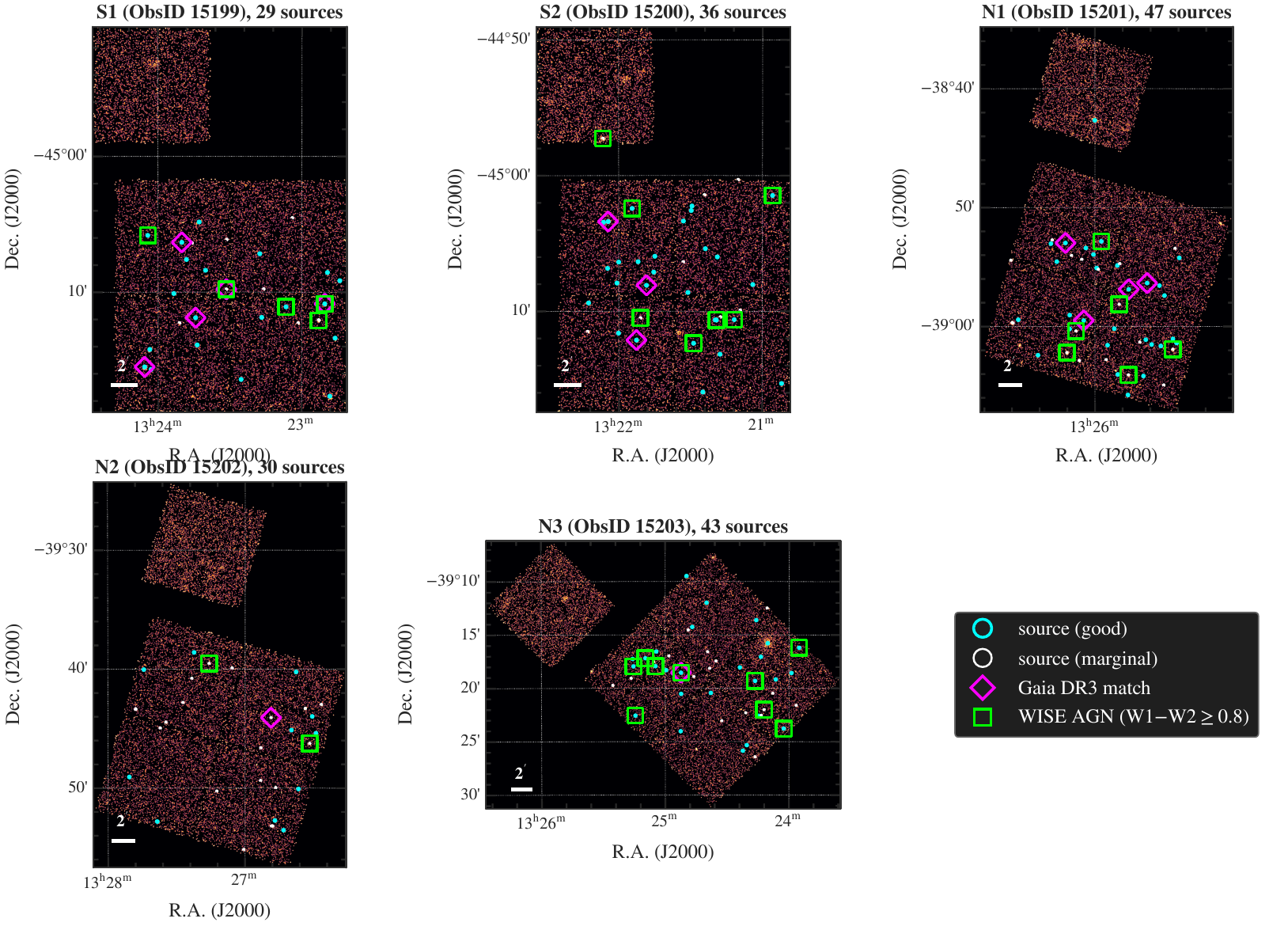}
  \caption{The five \chandra/ACIS-I giant-lobe fields (broad-band
  $0.5$--$7\keV$, adaptively smoothed, asinh stretch) with the vetted
  point sources overlaid: photometrically secure (``good'') sources as
  circles coloured by hardness ratio, ``marginal'' sources as squares,
  and the 14 Gaia~DR3 counterparts ringed in green.  Per-field source
  totals are given in each panel (185 in total).}
  \label{fig:fields}
\end{figure*}

\subsection{The point-source catalogue}
\label{sec:catalogue}

The final vetted catalogue of 185 sources is the basis for all that
follows; the sources are shown on the five fields in Fig.~\ref{fig:fields} and the first rows are listed in Table~\ref{tab:catalog}.  For each
source the full catalogue lists the IAU (CXOU) name, field and position,
the per-band net counts, the hardness ratio with its $68\%$ HPD interval
and the posterior aperture intensity $\lambda_S+\lambda_H$
(Appendix~\ref{app:hr}), the absorbed and unabsorbed $0.5$--$7\keV$
fluxes, the equivalent luminosity at $d=3.8$~Mpc, a photometric quality
flag, any Gaia~DR3 (with $G$) or AllWISE (with $W1-W2$) association, the
derived classification, positional, count-rate and flux uncertainties,
the PSF size, and the selection flags.  It is provided in machine-readable
form (\texttt{cena\_catalog.mrt} and \texttt{cena\_catalog.fits}) in the
arXiv ancillary material.

\begin{table*}
\caption{X-ray point-source catalogue of the Cen~A giant-lobe fields (first rows; full table of 185 sources available at the CDS).}
\label{tab:catalog}
\centering
\footnotesize
\begin{tabular}{rllrrrrrll}
\toprule
ID & CXOU name & Field & RA & Dec & $N_{\rm net}$ & HR & $\log L_{\rm X}$ & Qual. & Class \\
 & & & (deg) & (deg) & (ct) & & ($\log$\,erg\,s$^{-1}$) & & \\
\midrule
1 & CXOU J132523.5-385414 & N1 & 201.34802 & -38.90392 & 12 & 0.07 & 37.23 & good & uncl \\
2 & CXOU J132524.6-390123 & N1 & 201.35263 & -39.02330 & 59 & -0.16 & 37.96 & good & uncl \\
3 & CXOU J132525.3-385326 & N1 & 201.35555 & -38.89068 & 5 & 0.59 & 37.72 & marginal & uncl \\
4 & CXOU J132526.2-390103 & N1 & 201.35920 & -39.01754 & 30 & -0.53 & 37.67 & good & uncl \\
5 & CXOU J132526.2-390155 & N1 & 201.35944 & -39.03209 & 1 & -0.78 & 38.24 & marginal & AGN \\
6 & CXOU J132529.8-385724 & N1 & 201.37455 & -38.95676 & 25 & -0.24 & 37.54 & good & uncl \\
7 & CXOU J132530.1-390456 & N1 & 201.37583 & -39.08226 & 11 & -0.59 & 37.26 & marginal & uncl \\
8 & CXOU J132531.4-390138 & N1 & 201.38117 & -39.02743 & 18 & -0.57 & 37.50 & good & uncl \\
\bottomrule
\end{tabular}
\tablefoot{$N_{\rm net}$ is the broad-band ($0.5$--$7$\,keV) net count; HR$=(H-S)/(H+S)$ with $S=0.5$--$2$ and $H=2$--$7$\,keV (posterior median, Sect.~\ref{sec:detection} and Appendix~\ref{app:hr}); $\log L_{\rm X}$ assumes $d=3.8$\,Mpc. Qual.\ is the photometric quality flag (good/marginal; Sect.~\ref{sec:catalogue}). Class: AGN = WISE-selected ($W1-W2\geq0.8$), star = Gaia~DR3 counterpart that is not a WISE AGN, uncl.\ = unclassified. The full 185-row table, including fluxes and Gaia/AllWISE counterparts, is available in electronic form at the CDS (Sect.~\ref{sec:catalogue}).}
\end{table*}

\section{The lobe-field source population}
\label{sec:analysis}

Here we show, from hardness ratios, bright-source spectra and
completeness-corrected source counts, that the lobe-field population is
consistent with the foreground/background mixture expected at the
Galactic latitude of Centaurus~A ($b\approx19\fdg4$), and construct the
selection function $\Omega(S)$ used in Sect.~\ref{sec:constraints}.

\subsection{Hardness-ratio diagnostics}
\label{sec:hr}

\begin{figure}
  \centering
  \includegraphics[width=0.46\textwidth]{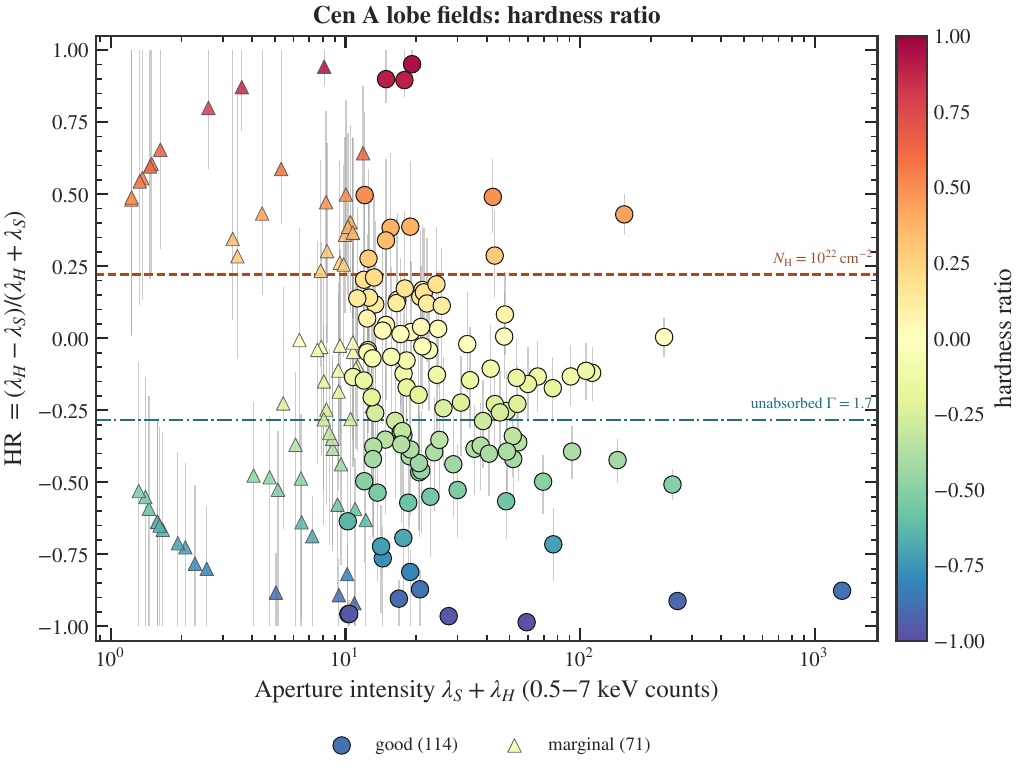}
  \caption{Bayesian hardness ratio $\mathrm{HR}=(H-S)/(H+S)$ (posterior
  median, $68\%$ HPD error bars; Appendix~\ref{app:hr}) versus the
  posterior median aperture intensity $\lambda_S+\lambda_H$
  ($0.5$--$7\keV$) for the 185 vetted sources.  Filled circles:
  photometrically secure (``good'') sources; open triangles: the fainter
  ``marginal'' tail.  Horizontal lines are the response-folded loci of
  Sect.~\ref{sec:hr}: unabsorbed $\Gamma=1.7$, $\mathrm{HR}_{0}=-0.28$
  (dot-dashed), and $\NH=10^{22}\,$cm$^{-2}$, $\mathrm{HR}_{22}=+0.22$
  (dashed).  The median is $\mathrm{HR}=-0.17$
  ($\NH\approx1.7\times10^{21}\,$cm$^{-2}$).}
  \label{fig:HR}
\end{figure}

We define the hardness ratio as
\begin{equation}
  \mathrm{HR} = \frac{H - S}{H + S},
  \label{eq:HR}
\end{equation}
which ranges from $-1$ (purely soft) to $+1$ (purely hard).  A single
two-band ratio does not uniquely determine a physical type --- HR depends
on photon index, absorbing column, redshift, off-axis response and, at
low counts, on Poisson noise --- so we use it only as a population
diagnostic.  Rather than adopt boundaries from the three-band
($\sim0.3$--$1$, $1$--$2$, $2$--$8\keV$) colour schemes of
\citet{Prestwich2003}, \citet{Sivakoff2008} and \citet{Burke2013}, which
are not defined on our $S=0.5$--$2.0$, $H=2.0$--$7.0\keV$ ratio, we fix
the bin edges from our own response: an absorbed power law is folded
through the ACIS-I ARF and RMF of a representative pointing in the same
two bands, and each edge is then a definite intrinsic column.  For
$\Gamma=1.7$ behind the Galactic column toward Cen~A
($\NH^{\rm Gal}=8.4\times10^{20}\,$cm$^{-2}$) this gives
\begin{itemize}
  \item $\mathrm{HR}_{0}=-0.28$ for an \emph{unabsorbed} power law
    ($-0.15$ for $\Gamma=1.4$, $-0.41$ for $\Gamma=2.0$), i.e.\ the
    unabsorbed locus is well below zero, not near it;
  \item $\mathrm{HR}_{22}=+0.22$ for one decade of intrinsic absorption,
    $\NH=10^{22}\,$cm$^{-2}$;
  \item for reference, $\mathrm{HR}=0$ already corresponds to
    $\NH\approx4.9\times10^{21}\,$cm$^{-2}$, and
    $\NH=10^{23}\,$cm$^{-2}$ saturates at $\mathrm{HR}=+0.97$, so no
    useful edge can be placed above $10^{22}\,$cm$^{-2}$ in this band
    pair.
\end{itemize}
We therefore bin the sample as \emph{consistent with no intrinsic
absorption} ($\mathrm{HR}<\mathrm{HR}_{0}$), \emph{mildly absorbed}
($\mathrm{HR}_{0}\leq\mathrm{HR}<\mathrm{HR}_{22}$, i.e.\ up to
$10^{22}\,$cm$^{-2}$) and \emph{absorbed}
($\mathrm{HR}\geq\mathrm{HR}_{22}$).  This is a
population split, not a per-source classifier: the edges sit within the
body of the distribution, and the $68\%$ HPD interval of $64\%$ of the
catalogue crosses at least one edge, so only the population fractions
should be read from these bins.  Figure~\ref{fig:HR} shows HR against
the posterior aperture intensity $\lambda_S+\lambda_H$, which, unlike
the net counts, is strictly positive and so defined for the faintest
detections (Appendix~\ref{app:hr}).

\subsection{The HR population}
\label{sec:hrpop}

The HR distribution of the 185 sources is summarised in
Table~\ref{tab:hrcensus} and has a median $\mathrm{HR}=-0.17$,
corresponding to $\NH\approx1.7\times10^{21}\,$cm$^{-2}$ at
$\Gamma=1.7$.  Measured against the response-folded edges of
Sect.~\ref{sec:hr}, $77$ sources ($42\%$) are consistent with no
absorption beyond the Galactic column, a further $71$ ($38\%$) are mildly
absorbed up to $10^{22}\,$cm$^{-2}$, and $37$ ($20\%$) lie above that
column.  This is the distribution expected of a population dominated by
background AGN: four-fifths of the sample is unabsorbed to mildly
absorbed, with a minority tail of more heavily absorbed objects, the
brightest of which (B5, $\NH\approx2.3\times10^{22}$~cm$^{-2}$;
Sect.~\ref{sec:specresults}) has a directly fitted column.

\begin{table}
\centering
\footnotesize
\caption{Hardness-ratio distribution of the 185 vetted sources (background-marginalised posterior medians, Appendix~\ref{app:hr}). Bin edges are response-folded for $\Gamma=1.7$ behind the Galactic column (Sect.~\ref{sec:hr}). A population split: the $68\%$ interval of $64\%$ of sources crosses an edge.}
\label{tab:hrcensus}
\begin{tabular}{@{}l@{\hspace{4pt}}c@{\hspace{4pt}}c@{\hspace{4pt}}p{2.9cm}@{}}
\toprule
Bin & HR & $N$ (\%) & $\NH$ \\
\midrule
No absorption & $<-0.28$ & 77 (42\%) & no excess over $N_{\rm H}^{\rm Gal}$ \\
Mild & $[-0.28,+0.22]$ & 71 (38\%) & $\leq 10^{22}$\,cm$^{-2}$ \\
Absorbed & $\geq+0.22$ & 37 (20\%) & $>10^{22}$\,cm$^{-2}$ \\
\bottomrule
\end{tabular}
\end{table}

A cross-match against Gaia~DR3 within $1.5''$ (comparable to the
quadrature sum of the Gaia and \chandra{} positional errors, the latter
dominated for these mostly on- to mid-axis sources by the
$\sim0.1$--$0.6''$ registration of Sect.~\ref{sec:reduction}) yields 14
optical counterparts.  From the
local Gaia surface density near Cen~A ($\approx1.2\times10^4$
deg$^{-2}$), a shifted-catalogue test gives $1.2$ expected chance
alignments ($8\%$ of the 14), so most are physical; individual faint,
large-off-axis matches are correspondingly less secure.  The
counterparts are preferentially the soft, bright sources: the two
brightest objects (HR $=-0.88$ and $-0.91$) coincide with $G\approx10$
stars whose Gaia~DR3 parallaxes ($6.94\pm0.02$ and $6.82\pm0.02$~mas,
i.e.\ $144$ and $147$~pc at $>350\sigma$, proper motions $29$--$31$~mas
yr$^{-1}$; Gaia positions propagated to the \chandra{} epoch) securely
identify them as nearby foreground stars.

We further cross-matched the catalogue against AllWISE
\citep{Wright2010} within $3''$ (AllWISE has a much broader PSF than
\chandra{}, so this is a generous radius).  Of the 185 sources, 93 have
a mid-IR counterpart; the local AllWISE surface density gives $8.4$
expected spurious associations (a $9\%$ false-match rate), so most are
real but individual matches are not uniformly secure.  Applying the
mid-IR colour criterion $W1-W2\geq0.8$~mag (Vega; the \citealt{Stern2012}
AGN wedge), 29 of the matched sources are \emph{WISE-selected AGN
candidates} and 64 have stellar/galaxy colours, independent
population-level support for the background-AGN-plus-foreground-star
interpretation.  The Gaia and WISE counterparts are marked in
Fig.~\ref{fig:fields}.

\subsection{Spectral parameters of the bright tail}
\label{sec:specresults}

For the eight brightest sources ($\geq100$ net broad-band counts;
labelled B1--B8 by decreasing counts, to avoid confusion with the
catalogue running IDs) we extracted source and background spectra with
\texttt{specextract} (source aperture enclosing $\approx90\%$ of the PSF,
a source-free surrounding annulus for the background, exposure-weighted
PSF-corrected ARF) over $0.5$--$7\keV$ and fitted them unbinned in
\xspec{}~v12.15.1 \citep{Arnaud1996} with the C-statistic
\citep{Cash1979} (source and background modelled jointly; W-statistic),
using \citet{Wilms2000} abundances.  The uniform baseline model is an
absorbed power law (\texttt{tbabs}$\times$\texttt{powerlaw}) with the
Galactic column frozen at $8.4\times10^{20}$~cm$^{-2}$ and thawed only
for the four sources with $>200$ counts (Table~\ref{tab:spectra}).  For
the five AGN candidates the median photon index is
$\widetilde{\Gamma}=1.4$ (range $0.4$--$2.6$; $\widetilde{\Gamma}
=1.6$ excluding the one absorbed source B5), consistent with background
AGN in deep \chandra{} surveys \citep{Tozzi2006,Ricci2017,Ueda2014}; we
do not include the three stellar sources in this statistic.

Three of the eight have bright Gaia~DR3 counterparts and stellar mid-infrared
colours ($W1-W2\approx0$), and are therefore foreground stars rather than
AGN: B1, B2 and B3, at $G=9.95$, $9.70$ and $12.38$.  Chance coincidence is
negligible at these magnitudes (fewer than $0.03$ matches expected across the
whole sample), so the identification does not rest on the X-ray spectrum ---
which, for B3 ($\mathrm{HR}=-0.51$, $\Gamma\approx2.2$), would not by itself
have separated it from a background AGN.

For the two softest sources (B1, B2; $\mathrm{HR}=-0.88,-0.91$) a power law is
not the appropriate model: the formal $\Gamma\approx5$--$6$ reflects a
soft, line-like spectrum, and the power-law fits leave coherent residuals
near $1\keV$.  A blackbody
(\texttt{tbabs}$\times$\texttt{bbodyrad}) converges to
$kT_{\rm bb}=0.23\pm0.01$ and $0.16\pm0.01\keV$, confirming soft, thermal
emission; a single-temperature optically-thin plasma (\texttt{apec}),
physically preferable for a corona, did not yield a stable fit in our
build, so we quote no plasma temperature and rest the foreground-star
identification on the Gaia counterparts and very soft HR rather than on
the X-ray model.

The hardest bright source (B5, $\mathrm{HR}=+0.43$) is formally flat
($\Gamma\approx0.4$) at the Galactic column, the classic signature of
unmodelled absorption.  With $\Gamma$ fixed at the canonical AGN value
$1.8$ and the column free (placed at $z=0$, as no redshift is known), we
obtain an intrinsic $\NH=(2.3\pm0.5)\times10^{22}$~cm$^{-2}$
(C-stat/dof $=105/112$, versus $125/112$ for the flat unabsorbed power
law at equal degrees of freedom, $\Delta C\approx-20$), i.e.\ a
\emph{moderately} obscured background AGN.  We report this as a $\Gamma$-fixed constraint because
the counts cannot break the $\Gamma$--$\NH$ degeneracy, and make no claim
of Compton-thick behaviour.

\begin{table*}
\centering
\caption{Absorbed power-law (\texttt{tbabs}$\times$\texttt{powerlaw}) fits to the 8 brightest sources ($\geq$100 net counts), labelled B1--B8 by decreasing counts (distinct from the catalogue running ID). Unbinned $0.5$--$7$\,keV spectra, source and background fitted jointly with the C-statistic (W-statistic), \citet{Wilms2000} abundances; C/dof is the fit statistic over degrees of freedom. $\Gamma$ and normalisation are free; $N_{\rm H}$ was thawed only for sources with $>$200 counts (otherwise frozen at the Galactic value). $F_{0.5-7}^{\rm unabs}$ is the unabsorbed $0.5$--$7$\,keV flux. Errors are 90\%.}
\label{tab:spectra}
\begin{tabular}{lccccccl}
\toprule
ID & Field & Net cts & $\Gamma$ & $N_{\rm H}$ & $F_{0.5-7}^{\rm unabs}$ & C/dof & Likely type \\
 & & (ct) & & ($10^{22}$\,cm$^{-2}$) & (erg\,cm$^{-2}$\,s$^{-1}$) & & \\
\midrule
B1 & S2 & 1305 & $5.20^{+0.39}_{-0.36}$ & $0.55$ & $1.7\times10^{-12}$ & 180/144 & foreground star\tablefootmark{b} \\
B2 & S1 & 254 & $6.38^{+1.21}_{-1.05}$ & $0.50$ & $3.6\times10^{-13}$ & 100/75 & foreground star\tablefootmark{b} \\
B3 & S1 & 247 & $2.20^{+0.72}_{-0.42}$ & $0.10$ & $3.1\times10^{-13}$ & 73/90 & foreground star\tablefootmark{d} \\
B4 & S1 & 225 & $1.38^{+0.44}_{-0.42}$ & $0.45$ & $3.3\times10^{-13}$ & 109/122 & AGN \\
B5 & N3 & 154 & $0.44^{+0.26}_{-0.26}$ & $0.084\tablefootmark{a}$ & $2.5\times10^{-13}$ & 125/112 & mod.\ obscured AGN\tablefootmark{c} \\
B6 & N3 & 144 & $2.64^{+0.45}_{-0.42}$ & $0.084\tablefootmark{a}$ & $2.4\times10^{-13}$ & 37/60 & AGN \\
B7 & S2 & 108 & $1.37^{+0.29}_{-0.29}$ & $0.084\tablefootmark{a}$ & $1.0\times10^{-13}$ & 65/87 & AGN \\
B8 & S1 & 106 & $1.79^{+0.52}_{-0.50}$ & $0.084\tablefootmark{a}$ & $1.7\times10^{-13}$ & 32/35 & AGN \\
\bottomrule
\end{tabular}
\tablefoot{\tablefoottext{a}{Frozen at the Galactic value, $8.4\times10^{20}$\,cm$^{-2}$.}
\tablefoottext{b}{Formal power-law $\Gamma$; a soft blackbody ($kT\approx0.16$--$0.23$\,keV) is a better description (Sect.~\ref{sec:specresults}).}
\tablefoottext{c}{Baseline (flat) fit shown; with $\Gamma=1.8$ fixed the column is free, giving $N_{\rm H}=(2.3\pm0.5)\times10^{22}$\,cm$^{-2}$ (C/dof $=105/112$).}
\tablefoottext{d}{Identified from its optical/infrared counterpart, not from the X-ray spectrum: a bright Gaia~DR3 source at $G=12.38$ (fewer than $0.03$ such matches are expected by chance across the whole sample) with a stellar mid-infrared colour, $W1-W2=-0.005$. Unlike B1 and B2 it is not exceptionally soft ($\mathrm{HR}=-0.51$), so its power-law index alone would not have distinguished it from an AGN.}}
\end{table*}

\subsection{Equivalent-luminosity distribution}
\label{sec:xlf}

\begin{figure}
  \centering
  \includegraphics[width=0.44\textwidth]{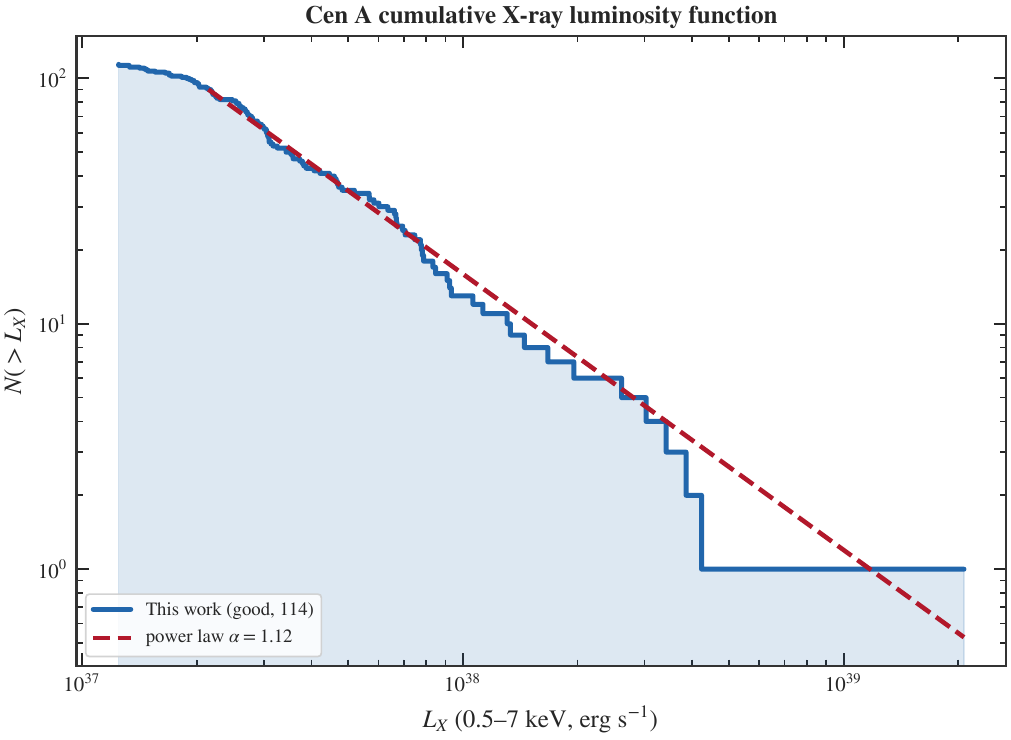}
  \caption{Cumulative distribution of the equivalent luminosity
  $N(>L_X)$ (flux at $d=3.8$~Mpc) of the photometrically secure
  (``good'') lobe-field sources; \emph{not} a physical XLF of a Cen~A
  population (Sect.~\ref{sec:xlf}).}
  \label{fig:xlf}
\end{figure}

Figure~\ref{fig:xlf} shows the cumulative distribution of the
\emph{equivalent} luminosity $L_X = 4\pi d^2 F$ that each good source
would have \emph{if} placed at $d=3.8$~Mpc (unabsorbed $0.5$--$7\keV$
fluxes).  As essentially all these sources are background AGN or
foreground stars, this is not a luminosity function of a Cen~A
population but the flux distribution in luminosity units; it fixes the
luminosity scale for a hypothetical lobe-associated population
(Sect.~\ref{sec:constraints}) and is incompleteness-limited at the faint
end.

\subsection{Source counts and the survey selection function}
\label{sec:lognlogs}

We compare the cumulative $0.5$--$2\keV$ $\log N$--$\log S$ with the
CXB, correcting for incompleteness with a per-pixel selection function on
the native $0.492''$ ACIS pixel grid.  Taking the \texttt{wavdetect}
\texttt{sigthresh}$=10^{-6}$ at face value as the per-position
false-alarm probability, the local detection threshold is the smallest
integer count that a background fluctuation reaches with probability
below it,
\begin{equation}
  C_{\rm lim}(B_{\rm ap}) = \min\{\,C\in\mathbb{N}:\
  P(N\!\geq\!C\,|\,B_{\rm ap}) \leq 10^{-6}\},
  \label{eq:clim}
\end{equation}
where $B_{\rm ap}$ is the local $0.5$--$2\keV$ background integrated over
an aperture of the $90\%$-encircled-energy radius read from the PSF map.
Equation~(\ref{eq:clim}) carries an explicit false-alarm rate and
follows the background level: $C_{\rm lim}$ is $3$ counts in the cleanest tenth of
the survey, $6$ at the median ($B_{\rm ap}=0.26$ counts), and $21$ in the
most contaminated tenth.  Completeness is then the probability that a
source of flux $S$ carries the aperture above that threshold,
\begin{equation}
  P_{\rm det}(S) = P\!\left(N \geq C_{\rm lim}\,\big|\,
  B_{\rm ap} + S\,t_{\rm exp}/K\right),
  \label{eq:pdet}
\end{equation}
with $t_{\rm exp}$ the soft-band exposure map --- which carries the
off-axis sensitivity, chip gaps and bad-pixel masking --- and $K$ the
count-to-flux conversion.  We derive $K$ from the instrument response
rather than from the detections, by folding the assumed absorbed
$\Gamma=1.7$ power law through each source's ARF; this gives
$K=1.57\times10^{-9}$~erg\,cm$^{-2}$, $7\%$ above the value obtained by
calibrating on the detected sources, and it is uniform between
observations to $1.4\%$.

\begin{figure}
  \centering
  \includegraphics[width=0.44\textwidth]{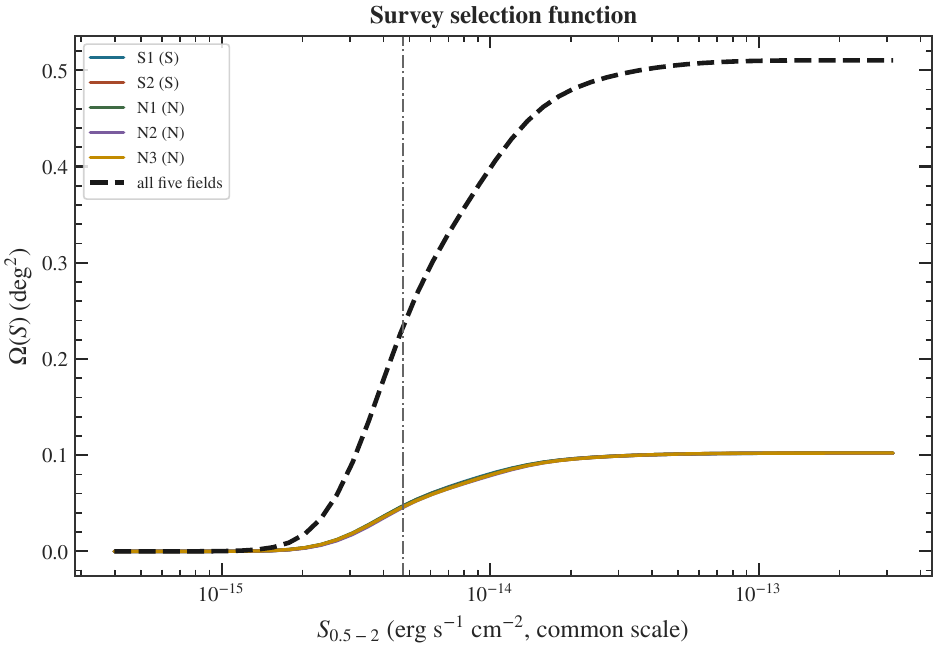}
  \caption{Selection function $\Omega(S)$ ($0.5$--$2\keV$, common flux
  scale) for the five \chandra{} pointings (S1, S2, N1, N2, N3), each
  contributing $\approx0.102$~deg$^2$ at near-identical depth, and their
  sum (dashed), which asymptotes to $0.51$~deg$^2$.  $\Omega$ is the
  completeness-weighted area of Eq.~(\ref{eq:omega}), including the
  broad-band quality cut.  The vertical line is the reporting threshold
  $S_{\rm ref}$.}
  \label{fig:omega}
\end{figure}

The analysis sample is not simply the detected sample: it is the
\emph{good} sample, which additionally requires broad-band
$N_{\rm net}\geq5$ and \texttt{SRC\_SIGNIFICANCE}$\geq3$.  In the data the
second cut is the binding one and it corresponds to
$N_{\rm net}\approx9$--$11$, so the selection function must model both.
Writing the broad-band aperture count as the soft count plus an
independent Poisson remainder, the probability that a source is detected
\emph{and} enters the good sample follows in closed form, and
\begin{equation}
  \Omega(S) = \sum_{\rm pix} P_{\rm sel}(S)\,A_{\rm pix}
  \label{eq:omega}
\end{equation}
is a completeness-weighted area rather than a step function (Fig.~\ref{fig:omega}).  The quality
cut costs little: it moves the flux at which half the footprint is
covered from $3.7$ to $4.3\times10^{-15}$~erg cm$^{-2}$ s$^{-1}$.

Two checks bound how far Eq.~(\ref{eq:clim}) can be trusted.  Internally,
every catalogued source must be detectable at its own position: under
Eq.~(\ref{eq:clim}) the median completeness at the positions of the 113
good sources is $0.98$ and $83\%$ have $P_{\rm det}>0.5$, whereas under
a conventional ``$n\sigma$'' aperture threshold
($C_{\rm lim}=\max[5,\,n^2/2+n\sqrt{B_{\rm ap}+(n/2)^2}]$ with $n=3.3$)
only $54\%$ of the same sources lay above their own limiting flux --- that
form is nearly background-independent, giving $11$ counts almost
everywhere, and is too conservative where the background is low and too
permissive where it is high.  Externally, the empirical \texttt{wavdetect}
sensitivity curves of \citet{Wright2023} imply a still lower threshold at
our background level; their tabulation is available to us only on-axis,
and applying an on-axis curve to sources spread over $0$--$10'$
over-corrects badly, so we do not use it.  Equation~(\ref{eq:clim}) is
therefore likely to remain somewhat conservative; the residual
uncertainty is carried explicitly in Sect.~\ref{sec:constraints}.

Our catalogue fluxes are \emph{absorbed} $0.5$--$2
\keV$ fluxes, attenuated by the Galactic column toward Cen~A
($8.4\times10^{20}\,$cm$^{-2}$), whereas \citet{Moretti2003} built their
counts with a conversion computed for $1.6\times10^{20}\,$cm$^{-2}$.  Folding both columns through our response gives a scale
factor of $1.198$ for $\Gamma=1.7$ (and $1.178$ for $\Gamma=1.4$, i.e.\ a
$1.7\%$ spectral systematic), which at the bright-end count slope
corresponds to a factor $1.39$ in $N(>S)$.  We apply it to both the
source fluxes and the selection function, so that all comparisons are
made on a single flux scale.

On that scale half the footprint is covered at
$S_{\rm ref}=4.8\times10^{-15}$~erg cm$^{-2}$ s$^{-1}$ and $90\%$ at
$1.5\times10^{-14}$; we report counts only above $S_{\rm ref}$, where the
$1/\Omega$ weights are bounded.  The total geometric area is $0.51$~deg$^2$.

\begin{figure}
  \centering
  \includegraphics[width=0.46\textwidth]{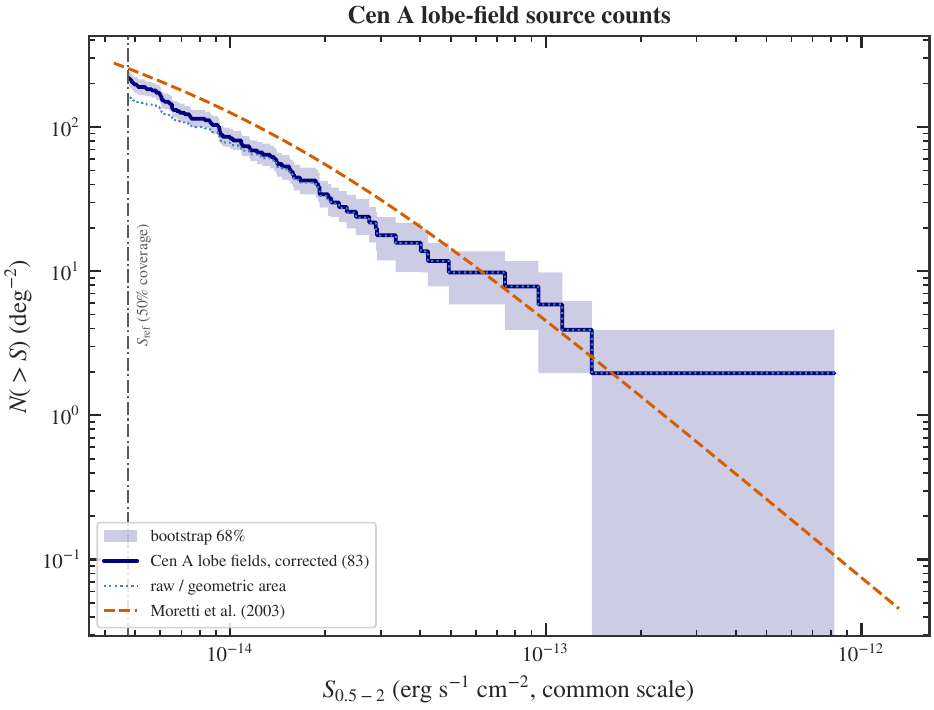}
  \caption{Completeness-corrected cumulative source counts $N(>S)$
  (deg$^{-2}$, $0.5$--$2\keV$) for the Cen~A lobe fields (navy), with the
  shaded band giving the $68\%$ bootstrap interval on the
  $1/\Omega$-weighted points, the uncorrected geometric-area counts
  (dotted), and the extragalactic background of \citet{Moretti2003} (red
  dashed).  The dot-dashed vertical line is the reporting threshold
  $S_{\rm ref}$, the flux at which the selection function covers half the
  footprint.  Fluxes are on the common scale of
  Sect.~\ref{sec:lognlogs}.}
  \label{fig:lognlogs}
\end{figure}

Figure~\ref{fig:lognlogs} compares the corrected counts with
\citet{Moretti2003}.  They run parallel to it and agree at the faint end
of our range ($0.81\pm0.10$ of the model at
$5\times10^{-15}$), falling to $0.66$ at $10^{-14}$ and $0.57$ at
$3\times10^{-14}$, where only a handful of sources contribute.  The
sources above $S_{\rm ref}$ account for $58\%$ of the soft CXB intensity.
The integrated comparison is made in Sect.~\ref{sec:constraints}.

\section{Discussion}
\label{sec:discussion}
\subsection{The nature of the lobe-field population}
\label{sec:interlopers}

\begin{figure}
  \centering
  \includegraphics[width=0.48\textwidth]{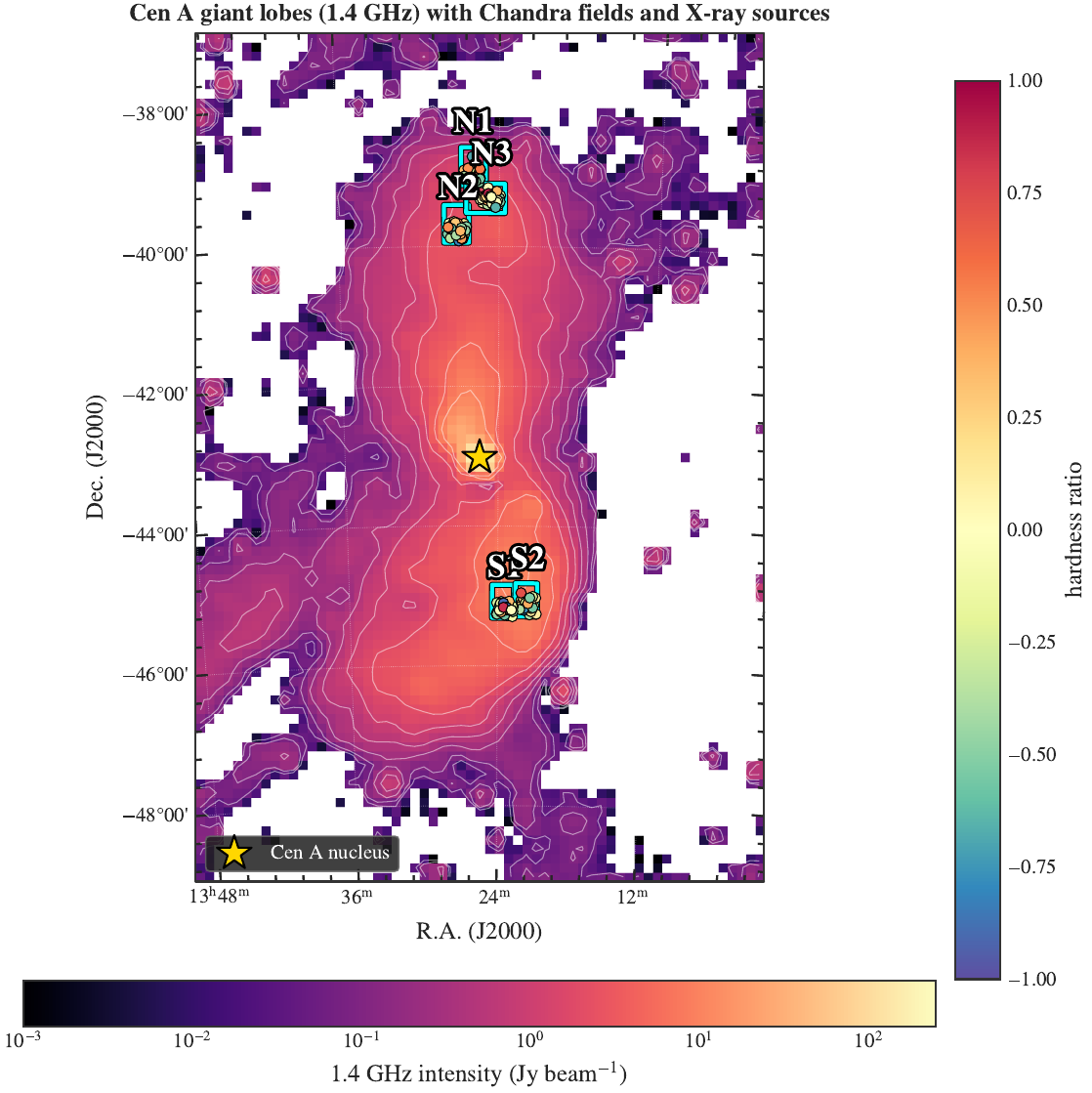}
  \caption{The five \chandra/ACIS-I fields (blue outlines, labelled
  S1--N3) and the 185 vetted X-ray point sources (coloured by hardness
  ratio) overlaid on the 1.4\,GHz total-intensity map of the Centaurus~A
  giant lobes \citep{OSullivan13}; the star marks the nucleus.}
  \label{fig:radio}
\end{figure}

\begin{figure}
  \centering
  \includegraphics[width=0.40\textwidth]{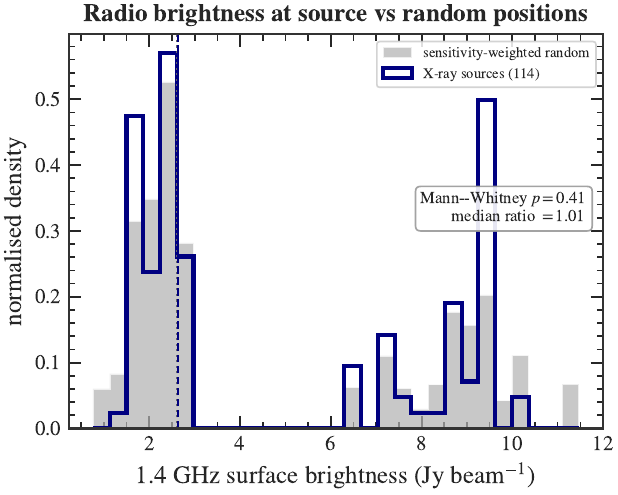}
  \caption{Distribution of the 1.4\,GHz surface brightness at the X-ray
  source positions (navy) and at sensitivity-weighted random positions
  within the ACIS footprints (grey); Mann--Whitney $p=0.41$
  (Sect.~\ref{sec:interlopers}).}
  \label{fig:radiohist}
\end{figure}

The fields lie $\gtrsim2^{\circ}$ ($\gtrsim130$~kpc in projection) from
the nucleus, far outside the LMXB-dominated inner galaxy.
Figure~\ref{fig:radio} shows them on the 1.4\,GHz total-intensity map of
the giant lobes.  We test for a concentration of sources on the radio
structure by comparing the 1.4\,GHz surface brightness at each
source position with a null distribution built from random positions
drawn within the ACIS footprints, weighted by the soft-band exposure map
so that the randoms follow the survey's spatially varying detection
probability.  The source and sensitivity-weighted random brightness
distributions are statistically indistinguishable
(Fig.~\ref{fig:radiohist}; median ratio $1.01$; Mann--Whitney $p=0.41$,
two-sample Kolmogorov--Smirnov $p=0.72$), and the
sources split $59{:}55$ between the radio-bright and radio-faint halves of
the footprint against $57{:}57$ expected ($p=0.71$).  There is thus no
preferential association with the brighter parts of the lobes, as
expected for a projected population.  Taken together, the
soft-dominated HR distribution (median $-0.17$), the AGN-like
bright-source indices ($\widetilde{\Gamma}\approx1.8$), the Gaia and WISE
identifications, and the count agreement indicate a population dominated
by (mostly unobscured) background AGN, with a minority of foreground
stars and a small absorbed-AGN tail.  We therefore test for an
\emph{excess} over this expected field population
(Sect.~\ref{sec:constraints}).

\subsection{A limit on a lobe-associated population}
\label{sec:constraints}

Of the 185 sources, 114 are ``good'', 113 of these have a positive
$0.5$--$2\keV$ flux, and 83 lie above $S_{\rm ref}$ and form the matched
sample (Table~\ref{tab:sampleflow}).  Twelve of the 113
have a Gaia~DR3 counterpart, but two also satisfy the mid-IR AGN colour
cut ($W1-W2\geq0.8$) and are faint in the optical ($G=17.2$ and $19.0$);
these two are counted as AGN.  Nine of the remaining foreground stars
survive the $S_{\rm ref}$ cut; the extragalactic \citet{Moretti2003}
counts do \emph{not} predict them, so they are modelled separately.  The
comparison uses one matched selection: the same band, area, flux
interval, quality cut (modelled inside $\Omega(S)$), star removal and
flux scale.

Above $S_{\rm ref}=4.8\times10^{-15}$~erg cm$^{-2}$ s$^{-1}$ the
$N_{\rm obs}=83$ sources include $N_{\rm star}=9$ foreground stars,
leaving $74$ AGN candidates.  Folding the
\citet{Moretti2003} differential counts through $\Omega(S)$ predicts
$N_{\rm exp}=101.4$.  The observed count is therefore \emph{below} the
background expectation, by $27$ sources or $27\%$:
\begin{equation}
  N_{\rm AGN} - N_{\rm exp} = 74 - 101.4 = -27.4 .
  \label{eq:diff}
\end{equation}
The equivalent luminosity at $S_{\rm ref}$ and $d=3.8$~Mpc is
$6.9\times10^{36}\erg$; half the footprint is sensitive to that
luminosity, rising to $90\%$ at $2.2\times10^{37}\erg$.

Taken as pure Poisson noise this shortfall would be significant
($P=0.005$), but that neglects the normalisation of the count model,
which is the dominant uncertainty for a survey this shallow.  Modern
soft-band determinations span $\approx15\%$ in normalisation; the
field-to-field scatter of our own five pointings (ratios $0.70$, $0.75$,
$0.90$, $0.40$, $0.96$; Table~\ref{tab:perfield}) has an rms of $0.21$
against $0.22$ expected from Poisson noise alone, so we detect no
additional cosmic-variance term and adopt $\sigma_{\rm bkg}=15\%$.
Including it, the shortfall is $-1.5\sigma$ and is not significant.
Table~\ref{tab:cxbmodels} shows that varying the normalisation and the
faint-end slope over the range spanned by modern determinations moves the
difference between $-0.8\sigma$ and $-2.1\sigma$: the sign is robust, the
significance is not.

We do not interpret this as a physical underdensity of background AGN
behind the lobes.  A $27\%$ shortfall is comparable to what remains
uncertain in the selection function itself: Eq.~(\ref{eq:clim}) is a
Poisson aperture criterion standing in for \texttt{wavdetect}'s wavelet
correlation, and the comparison with \citet{Wright2023} in
Sect.~\ref{sec:lognlogs} indicates it errs on the conservative side, which
biases $\Omega$ --- and hence $N_{\rm exp}$ --- high.  Scaling the detection
threshold by $\pm30\%$ moves $N_{\rm exp}$ by only $^{+6}_{-9}$, so the
threshold model alone does not account for it, but a calibrated
completeness curve could.  For the question asked here the direction is
what matters: there is no surplus over the field expectation.

To set a limit we model the observed count as
$N_{\rm obs}\sim\mathrm{Poisson}(N_{\rm CXB}(1+\delta)+N_{\rm star}+
N_{\rm lobe})$ with $N_{\rm lobe}\geq0$ and marginalise the background
normalisation $\delta\sim\mathcal{N}(0,\sigma_{\rm bkg})$.  The $90\%$
posterior upper limit on a lobe-associated population above $S_{\rm ref}$
is
\begin{equation}
  N_{\rm lobe} < 18 \quad (90\%,\ \mathrm{marginalised}),
  \label{eq:ul}
\end{equation}
tightening to $<8$ if the CXB normalisation is held fixed and relaxing to
$<38$ under a conservative $30\%$ normalisation uncertainty; across the
count models of Table~\ref{tab:cxbmodels} it stays within $16$--$21$.
This limit is tighter than for an exactly matched field because the
observed count already sits below the expectation; it is conditional on
the CXB prediction being correct.  Equation~(\ref{eq:clim}) remains an
analytic stand-in for \texttt{wavdetect}; an injection--recovery run
(simulated sources across the real backgrounds and off-axis angles,
re-detected with the same chain) would replace it with a calibrated curve
and is the change most likely to move $N_{\rm exp}$.  Until then we quote
the conservative $N_{\rm lobe}<38$ as the headline limit.

\begin{table}
\centering
\scriptsize
\caption{Per-field soft-band ($0.5$--$2\keV$) counts above the reporting threshold $S_{\rm ref}$ (Sect.~\ref{sec:lognlogs}).  $N_{\rm obs}$ is the good sample, $N_{\rm star}$ the foreground stars within it, $N_{\rm exp}$ the \citet{Moretti2003} counts folded through that field's selection function.  $S_{50}$ is the flux at which the field's selection function covers half its area. The per-field $N_{\rm exp}$ are computed on each field's own binned selection function and sum to $99.6$, within $2\%$ of the $101.4$ obtained from the combined function used in Sect.~\ref{sec:constraints}.}
\label{tab:perfield}
\begin{tabular}{@{}l@{\hspace{4pt}}l@{\hspace{4pt}}c@{\hspace{4pt}}c@{\hspace{4pt}}c@{\hspace{4pt}}c@{\hspace{4pt}}c@{\hspace{4pt}}c@{}}
\toprule
Field & Lobe & Area & $N_{\rm obs}$ & $N_{\rm star}$ & $N_{\rm AGN}$ & $N_{\rm exp}$ & $S_{50}$ \\
 & & (deg$^2$) & & & & & ($10^{-15}$) \\
\midrule
S1 & S & 0.102 & 17 & 3 & 14 & 20.1 & 5.06 \\
S2 & S & 0.102 & 18 & 3 & 15 & 19.9 & 5.11 \\
N1 & N & 0.102 & 21 & 3 & 18 & 20.0 & 5.06 \\
N2 & N & 0.102 & 8 & 0 & 8 & 19.8 & 5.18 \\
N3 & N & 0.102 & 19 & 0 & 19 & 19.8 & 5.16 \\
\midrule
All & --- & 0.510 & 83 & 9 & 74 & 99.6 & --- \\
\bottomrule
\end{tabular}
\end{table}

\begin{table}
\centering
\footnotesize
\caption{Sensitivity of the count comparison to the assumed CXB soft-band count model, all evaluated above $S_{\rm ref}$ on the common flux scale (Sect.~\ref{sec:lognlogs}).  $N_{\rm exp}$ is the model folded through the selection function; the difference is $N_{\rm AGN}-N_{\rm exp}$ with $N_{\rm AGN}=74$; $z$ includes the 15\% systematic on $N_{\rm exp}$; UL is the $90\%$ marginalised limit on $N_{\rm lobe}$.}
\label{tab:cxbmodels}
\begin{tabular}{@{}lrrrc@{}}
\toprule
CXB count model & $N_{\rm exp}$ & difference & $z$ & UL$_{90}$ \\
\midrule
Moretti (2003), fiducial & 101.4 & $-27.4$ & $-1.5$ & $<18$ \\
normalisation $-15\%$ & 86.2 & $-12.2$ & $-0.8$ & $<21$ \\
normalisation $+15\%$ & 116.6 & $-42.6$ & $-2.1$ & $<16$ \\
faint slope $a_2{=}0.70$ & 109.3 & $-35.3$ & $-1.8$ & $<17$ \\
faint slope $a_2{=}0.50$ & 93.9 & $-19.9$ & $-1.2$ & $<19$ \\
\bottomrule
\end{tabular}
\end{table}

The result is not driven by a single field (Table~\ref{tab:perfield}):
the per-field ratios show no north--south asymmetry and Poisson-level
scatter.  N2 is lowest ($0.40$), but its own selection function is
comparable to the others, so this is a local fluctuation rather than a
failure of the sensitivity model.  Any population physically associated
with the lobes
--- low-mass X-ray binaries in stripped or halo stellar material
\citep{Gilfanov2004,Kim2004}, high-mass X-ray binaries requiring in-situ
star formation \citep{Grimm2003,Mineo2012}, or relic compact jet/lobe
structures --- is therefore limited to a detected excess of
$N_{\rm lobe}<38$ above $6.9\times10^{36}\erg$ over the half of the
footprint covered at $S_{\rm ref}$.  We do
not convert this into stellar-mass or star-formation-rate limits: the
constraint is a soft-band detected-excess count, whereas the LMXB
\citep{Gilfanov2004} and HMXB \citep{Mineo2012} luminosity functions are
calibrated in a broader band, and an intrinsic X-ray-binary population
would further require a spatial and luminosity-function model that these
shallow data do not constrain.  We find no independent evidence (e.g.\ of
recent star formation) for any such component.

\subsection{Implications for diffuse X-ray and IC measurements}
\label{sec:IC}

Because the completeness-corrected source counts match the extragalactic
background (Sect.~\ref{sec:lognlogs}), with no significant excess
(Sect.~\ref{sec:constraints}), the lobe-projected point sources are
consistent with the CXB seen through the lobe fields.  The resolved
sources account for $\approx50\%$ of the $0.5$--$2\keV$ CXB surface
brightness ($3.8\times10^{-12}$ of $7.5\times10^{-12}$~erg cm$^{-2}$
s$^{-1}$ deg$^{-2}$; \citealt{Moretti2003}); the remaining unresolved
background is the standard CXB present in any blank field.

The field inferred from the synchrotron-to-IC ratio scales as
$B\propto(S_{\rm sync}/S_{\rm IC})^{1/(1+\alpha)}$ at fixed CMB energy
density (reducing to the often-quoted $1/2$ only for a synchrotron
spectral index $\alpha=1$), so
any point-source flux $\Delta F$ that is \emph{not} removed before the IC
measurement biases the inferred field by
\begin{equation}
  \frac{\Delta B}{B} \simeq -\frac{1}{1+\alpha}\,\frac{\Delta F}{F_{\rm IC}},
  \label{eq:bbias}
\end{equation}
with $F_{\rm IC}$ the diffuse IC lobe flux in the same band and aperture.
This correction is \emph{not} automatic: background subtraction removes
resolved point sources only if the IC analysis applies the same masking
and source treatment, including flux leaking outside finite source masks
and fluctuations of the unresolved CXB.  Our result bounds the
\emph{number} of excess sources ($N_{\rm lobe}<38$); converting it to
$\Delta F$, and thence to $\Delta B/B$ via Eq.~(\ref{eq:bbias}), requires
an assumed luminosity function for the putative lobe population and
$F_{\rm IC}$ in a matched band and aperture, which is poorly constrained
because the diffuse lobe emission blends thermal, synchrotron and IC
components \citep{Stawarz2013}.  We therefore quote no single
$\Delta B/B$: these data reveal no lobe-associated point-source component
beyond the standard CXB, so no additional point-source correction to the
$\mu$G-level IC field of the Cen~A lobes
\citep{Hardcastle2009,Yang2012} is indicated, provided the IC analysis
subtracts resolved sources and the blank-field CXB as usual.

\subsection{Future prospects}
\label{sec:future}

Factor-of-several deeper \chandra{} exposures of the lobe fields would
push the detection floor below $10^{36}\erg$ and allow a genuine
population synthesis, although at large off-axis angles the gain is
limited by the broad PSF and the particle background.  A quantitative correlation of the source
density with the radio-lobe surface brightness, and incorporation of the
deep ${\sim}\,800$~ks Cen~A VLP data over the inner galaxy to connect this
census to the LMXB population, are natural next steps.  Because of its
$\sim1'$ angular resolution and small field, \textit{XRISM}/Resolve is
not suited to resolving these faint point sources, but it can characterise
the integrated diffuse (thermal/IC) spectrum; the \textit{Athena} Wide
Field Imager, with its high throughput and good angular resolution, is the
natural instrument for extending this point-source census.

\section{Conclusions}
\label{sec:conclusions}

\begin{enumerate}
\item From five dedicated \chandra/ACIS-I pointings we build a
  catalogue of 185 vetted X-ray point sources projected against the giant
  radio lobes of Centaurus~A (arXiv ancillary material) and test for a
  population in excess of the expected field population.
\item The sample is consistent with a background-AGN-plus-foreground-star
  mixture: a soft-dominated HR distribution (median $\mathrm{HR}=-0.17$;
  $80\%$ absorbed by less than $10^{22}\,$cm$^{-2}$), an AGN median bright-source index
  $\widetilde{\Gamma}\approx1.6$ with one moderately obscured AGN
  ($\NH\approx2.3\times10^{22}$~cm$^{-2}$), 14 Gaia~DR3 counterparts
  ($\approx1.2$ expected by chance; the two brightest are foreground
  stars at $144$ and $147$~pc), 29 WISE-selected AGN candidates, and a
  completeness-corrected $\log N$--$\log S$ that tracks
  \citet{Moretti2003} to within $20$--$40\%$ over the range we report.
\item Above $S_{\rm ref}$, where half the footprint is covered, the
  good sample holds $N_{\rm obs}=83$ sources, $9$ of them foreground
  stars, leaving $74$ AGN candidates against $N_{\rm exp}=101$ from
  \citet{Moretti2003} folded through the selection function: $27\%$
  \emph{below} the expectation, $-1.5\sigma$ including the $15\%$
  normalisation uncertainty, and no excess.  A star-accounted Poisson
  likelihood marginalised over that normalisation gives
  $N_{\rm lobe}<18$ ($90\%$), or $<38$ for a conservative $30\%$
  normalisation uncertainty, above $6.9\times10^{36}\erg$.  This limits
  the detected excess in the surveyed footprint, not stellar mass or star
  formation (Sect.~\ref{sec:constraints}).
\item We find no lobe-associated point-source component beyond the
  standard CXB in the observed fields, so resolved point sources do not
  contaminate diffuse-emission (including inverse-Compton) measurements of
  the lobes beyond the standard CXB, provided resolved sources and the
  blank-field CXB are subtracted as usual.  A quantitative $\Delta B/B$
  bound would require an assumed lobe luminosity function and a
  matched-aperture $F_{\rm IC}$, which these data do not constrain.
\item The released catalogue provides a baseline for deeper \chandra{},
  \textit{Athena} and diffuse-emission \textit{XRISM} observations.
\end{enumerate}

\begin{acknowledgements}
K.B.\ acknowledges support from the PETC Overseas Scholarship, Government
of Karnataka, India (award DSTW/ED-1/CR-4/2016-17; 2016--2021).  K.B.'s
postdoctoral fellowship at National Tsing Hua University
(Aug.\ 2022--Jul.\ 2023), and H.-K.C.'s contribution, were supported by
MOST grant 111-2112-M-007-019 (NTHU project 111B0034I9; MOST is now the
National Science and Technology Council, NSTC).  {\L}.S.\ and V.M.\
acknowledge support from Polish National Science Centre (NCN) grant
2016/22/E/ST9/00061.

We thank Vinay L.\ Kashyap for valuable technical comments on the Bayesian
hardness-ratio analysis, background treatment, and survey-sensitivity
calculations that substantially improved the manuscript.

This work is based on archival observations obtained with the
\chandra{} X-ray Observatory under the programme \emph{Studying Giant
Filamentary Lobes of Centaurus~A} (PI: Y.~T.~Tanaka; ObsIDs
15199--15203, Table~\ref{tab:obslog}), and has made use of the
\chandra{} Data Archive and of software provided by the \chandra{}
X-ray Center (CXC), which is operated by the Smithsonian Astrophysical
Observatory for and on behalf of the National Aeronautics and Space
Administration (NASA) under contract NAS8-03060.
The 1.4~GHz total-intensity map of the giant lobes is from the Parkes observations of \citet{OSullivan13}; the Parkes radio telescope is part of the Australia Telescope National Facility, funded by the Commonwealth of Australia for operation as a National Facility managed by CSIRO.

This work has made use of data from the European Space Agency (ESA)
mission {\it Gaia} (\url{https://www.cosmos.esa.int/gaia}), processed by
the {\it Gaia} Data Processing and Analysis Consortium (DPAC,
\url{https://www.cosmos.esa.int/web/gaia/dpac/consortium}).  Funding for
the DPAC has been provided by national institutions, in particular the
institutions participating in the {\it Gaia} Multilateral Agreement.
This publication also makes use of data products from the Wide-field
Infrared Survey Explorer, which is a joint project of the University of
California, Los Angeles, and the Jet Propulsion
Laboratory/California Institute of Technology, funded by NASA.

Data reduction used \textsc{CIAO}~4.18 and \textsc{CALDB}~4.12.4
\citep{Fruscione2006}; spectral fitting used \xspec{}~v12.15.1
\citep{Arnaud1996} and \textsc{Sherpa} \citep{Freeman2001}.  This work
made use of \texttt{astropy} \citep{Astropy2013,Astropy2018,Astropy2022},
\texttt{astroquery} \citep{Astroquery2019}, \texttt{numpy}
\citep{NumPy2020}, \texttt{scipy} \citep{SciPy2020} and
\texttt{matplotlib} \citep{Matplotlib2007}, and of NASA's Astrophysics
Data System Bibliographic Services.
\end{acknowledgements}

\bibliographystyle{aa}
\bibliography{references_paper1}

\begingroup\let\clearpage\relax
\begin{appendix}
\section{Aperture intensities and hardness ratios}
\label{app:hr}

\subsection{Photometric model}
\label{app:hr:model}

Aperture photometry with \texttt{srcflux} supplies, for each source and
each band, the gross counts $C$ in the source aperture of area
$\Omega_s$, the counts $B$ in the surrounding background aperture of area
$\Omega_b$ ($\Omega_b/\Omega_s = 22$ in the median, and $\geq14$
throughout), and the fractions $f$ and $g$ of the local PSF enclosed by
the two apertures.  Following \citet{Primini2014}, the two aperture
counts are independent Poisson variables whose means \emph{both} contain
a source and a background term,
\begin{equation}
  \mu_s = \lambda + \Omega_s\,b, \qquad
  \mu_b = \gamma\,\lambda + \Omega_b\,b, \qquad
  \gamma \equiv g/f,
  \label{eq:apmeans}
\end{equation}
where $b$ is the background surface density and $\lambda$ is the expected
number of \emph{source} counts inside the source aperture.

Equation~(\ref{eq:apmeans}) is written in the aperture frame, i.e.\
$\lambda$ is an aperture quantity rather than the PSF-corrected source
intensity, and the enclosed-PSF fractions enter only through their ratio
$\gamma$.  This is deliberate, and it is what makes the measured HR
directly comparable to the response-folded HR loci of Sect.~\ref{sec:hr}:
the per-source ARFs were produced with \texttt{specextract}
(\texttt{correctpsf=no}) followed by \texttt{arfcorr}, so they already
carry the energy-dependent enclosed-PSF fraction --- falling from $0.90$
at $0.6\keV$ to $0.76$ at $7\keV$ for a typical aperture --- and a model
folded through them therefore predicts the counts \emph{collected in the
aperture}.  Measuring HR from PSF-corrected intensities while comparing it
to loci folded through aperture-corrected ARFs would mix the two frames;
the mismatch is not negligible.  For the 165 sources whose apertures
enclose more than half of the local PSF the band ratio is
$f_S/f_H = 1.069$ (full range $1.063$--$1.099$), which alone would shift
the sample median HR by $+0.023$ and move five sources across the soft
boundary (Table~\ref{tab:hrvariants}).  The remaining 20 sources are all
marginal detections of $1$--$5$ counts whose \texttt{wavdetect} ellipses
enclose only $0.2$--$12\%$ of the local PSF; in the aperture frame they
require no PSF correction at all, and they enter only through
$\gamma>1$ for four of them, i.e.\ apertures that leak more source flux
into the background region than they retain.

\subsection{Posterior for the band intensity}
\label{app:hr:post}

With Jeffreys priors on both intensities, $p(\lambda)\propto
\lambda^{-1/2}$ and $p(b)\propto b^{-1/2}$, the posterior for the band
intensity marginalised over the background is
\begin{equation}
  p(\lambda \,|\, C,B) \;\propto\; \lambda^{-1/2}
  \int_{0}^{\infty}\!\mathrm{d}b\;\, b^{-1/2}\,
  \mu_s^{\,C}\,e^{-\mu_s}\;\mu_b^{\,B}\,e^{-\mu_b}.
  \label{eq:lampost}
\end{equation}
This is \emph{not} a gamma density.  For the conjugate gamma-prior family
of \citet{Primini2014} with non-informative parameters
($\alpha_s=\alpha_b=1$, $\beta_s=\beta_b=0$) the integral can be done in
closed form (their Eq.~16), which for the apertures of
Eq.~(\ref{eq:apmeans}) reads
\begin{equation}
  p(\lambda \,|\, C,B) \;\propto\; e^{-(1+\gamma)\lambda}
  \sum_{k=0}^{C}\;\sum_{j=0}^{B}\; w_{kj}\,\lambda^{\,k+j},
  \label{eq:pk14}
\end{equation}
\begin{equation}
  w_{kj} = \frac{\gamma^{\,j}\;\Omega_s^{\,C-k}\;\Omega_b^{\,B-j}}
                {k!\,j!\,(C\!-\!k)!\,(B\!-\!j)!}\;
           \frac{\Gamma(C\!-\!k\!+\!B\!-\!j\!+\!1)}
                {(\Omega_s+\Omega_b)^{C-k+B-j+1}} .
  \label{eq:pk14w}
\end{equation}
Each term is a gamma density of shape $k+j+1$ and rate $1+\gamma$, so the
posterior is a finite mixture of $(C+1)(B+1)$ gamma components --- one for
every way of attributing the counts in the two apertures to source or to
background --- and collapses to a single gamma density only in the
limit of vanishing background.  With the Jeffreys priors adopted here the
mixture shapes become $k+j+\tfrac12$ and the summation limits are no
longer integers, so we evaluate Eq.~(\ref{eq:lampost}) numerically, on a
$\sqrt{\lambda}$ grid that absorbs the integrable $\lambda^{-1/2}$
singularity at the origin exactly, with the background marginalised by
quadrature.  Run with flat priors, that calculation reproduces
Eq.~(\ref{eq:pk14}) to better than $4\times10^{-4}$ in log density and to
a relative $10^{-6}$ in the posterior median, tested on 30 real
band--source combinations spanning the catalogue from $C=0$ to $C=818$.

\begin{table}
\centering
\scriptsize
\caption{Hardness-ratio census under the adopted posterior and four
alternatives.  Shifts $\Delta$HR are per source, relative to the first
row.  The census columns use the response-folded bins of
Sect.~\ref{sec:hr}.  The last row mixes frames
(Appendix~\ref{app:hr:model}) and is listed only to size that effect; it
is not used.}
\label{tab:hrvariants}
\begin{tabular}{@{}l@{\hspace{5pt}}c@{\hspace{5pt}}c@{\hspace{5pt}}c@{\hspace{5pt}}c@{}}
\toprule
Estimator & med. & med. & max & none/mild/abs.\ \\
          & HR   & $|\Delta$HR$|$ & $|\Delta$HR$|$ & $N$ \\
\midrule
Fixed bkg.\tablefootmark{a}      & $-0.173$ & ---     & ---     & 72/76/37 \\
\textbf{Adopted}\tablefootmark{b} & $-0.173$ & $0.017$ & $0.51$  & 77/71/37 \\
$\gamma=0$\tablefootmark{c}      & $-0.173$ & $0.007$ & $0.23$  & 74/74/37 \\
Flat priors\tablefootmark{d}     & $-0.162$ & $0.016$ & $0.18$  & 72/77/36 \\
PSF-corr.\ frame\tablefootmark{e} & $-0.145$ & $0.029$ & $0.54$ & 75/71/39 \\
\bottomrule
\end{tabular}
\tablefoot{
\tablefoottext{a}{$\lambda+b$ drawn from a gamma density truncated at
$b=$ gross $-$ net.}
\tablefoottext{b}{Eq.~(\ref{eq:lampost}): background marginalised,
$\gamma=g/f$ retained, Jeffreys priors.}
\tablefoottext{c}{As adopted, but with the background aperture assumed
free of source counts.}
\tablefoottext{d}{As adopted, but with the flat priors of
Eq.~(\ref{eq:pk14}).}
\tablefoottext{e}{As adopted, but with $\lambda$ taken as the
PSF-corrected source intensity, which is inconsistent with the
aperture-corrected ARFs used for the HR loci of Sect.~\ref{sec:hr}.}}
\end{table}

\subsection{Hardness ratio}
\label{app:hr:hr}

The two bands are independent, so their posteriors are sampled
independently ($6\times10^{4}$ draws each) and combined into
\begin{equation}
  \mathrm{HR} = \frac{\lambda_H-\lambda_S}{\lambda_H+\lambda_S},
  \label{eq:hrlam}
\end{equation}
which is bounded to $[-1,+1]$ by construction because both intensities are
non-negative.  The HPD interval is the shortest interval of its
probability content by definition, so it is narrower than the equal-tail
interval for any asymmetric posterior and equal to it only for a
symmetric one; none of our 185 posteriors is symmetric, which is why the
comparison quoted in Sect.~\ref{sec:detection} holds for every source
rather than for most of them.  The soft band is formed by combining the \textsc{CIAO}
soft and medium bands into a single $0.5$--$2.0\keV$ band before the
posterior is constructed, with $f$ and $g$ taken as the gross-count
weighted means of the two sub-bands; treating them as two independent
bands and adding the intensities would add half a prior count to $S$ and
shift the sample median HR by $-0.03$.  We quote the posterior median and
the $68\%$ HPD interval (Sect.~\ref{sec:detection}), and the same draws
give the aperture intensity $\lambda_S+\lambda_H$ used as the abscissa of
Fig.~\ref{fig:HR}; its median exceeds the broad-band net counts by $5\%$,
the difference being the background that the net counts subtract but the
posterior marginalises.  Our implementation is not the BEHR code of
\citet{Park2006}: it follows the same bounded-estimator idea but
evaluates Eq.~(\ref{eq:lampost}) directly, with the background treatment
of \citet{Primini2014}.

\subsection{How much the modelling choices matter}
\label{app:hr:variants}

Table~\ref{tab:hrvariants} recomputes the catalogue under the adopted
posterior and under four alternatives.  The first row is a simpler
estimator that draws $\lambda+b$ from a gamma density truncated at a
background held fixed at gross minus net; our code reproduces its original
output to a median $0.0015$ (maximum $0.0138$) and its half-widths to
$0.0013$.  Marginalising the background properly shifts the sample median
HR by $-0.003$ and the individual values by a median $0.017$; the shift
stays below $0.33$ for every source with a well-behaved aperture and
reaches $0.51$ for one $2$-count marginal source.  Five sources move
between the unabsorbed and mildly absorbed bins of Sect.~\ref{sec:hr};
the absorbed bin is unchanged.  The choice of prior (Jeffreys
versus flat) moves the sample median by a comparable $+0.011$, and
retaining the source's leakage into the background aperture accounts for
about half of the total shift.  None of these changes affects any
conclusion of Sect.~\ref{sec:hrpop}: the distribution remains
soft-dominated with a median near $-0.17$, and the fraction consistent
with no absorption beyond the Galactic column stays within $39$--$42\%$.

\end{appendix}
\endgroup

\end{document}